# Improving the Accuracy of Temperature Measurement on TEM sample using Plasmon Energy Expansion Thermometry (PEET): Addressing Sample Thickness Effects

Yi-Chieh Yang[1,&], Luca Serafini[2,3,&], Nicolas Gauquelin[2,3], Johan Verbeeck[2,3], Joerg R. Jinschek[1,*]

[1] National Centre for Nano Fabrication and Characterization (DTU Nanolab), Technical University of Denmark (DTU), Kgs. Lyngby, Denmark.

[2] Electron Microscopy for Materials Science (EMAT), University of Antwerp, Antwerp, Belgium.

[3] NANOlab Center of Excellence, University of Antwerp, Antwerp, Belgium.

[&] These authors contributed equally
[*] Corresponding author: jojin@dtu.dk

## *Abstract*

Advances in analytical scanning transmission electron microscopy (STEM) with microelectronic mechanical systems (MEMS) based microheaters have enabled in-situ materials' characterization at the nanometer scale at elevated temperature. In addition to resolving the structural information at elevated temperatures, detailed knowledge of the local temperature distribution inside the sample is essential to reveal thermally induced phenomena and processes. Here, we investigate the accuracy of plasmon energy expansion thermometry (PEET) as a method to map the local temperature in a tungsten (W) lamella in a range between room temperature and 700 ºC. In particular, we address the influence of sample thickness in the range of a typical electron-transparent TEM sample (from 30 nm to 70 nm) on the temperature-dependent plasmon energy. The shift in plasmon energy, used to determine the local sample temperature, is not only temperature-dependent, but in case of W also thickness-dependent in sample thicknesses below approximately 60 nm. The results highlight the importance of considering sample thickness (and especially thickness variations) when analyzing the local bulk plasmon energy for temperature measurement using PEET. However, in case of W, an increasing beam broadening (FWHM) of the bulk plasmon peak with decreasing sample thickness can be used to improve the accuracy of PEET in TEM lamellae with varying sample thickness.

*Keywords: bulk plasmon energy, tungsten, FIB, STEM-EELS, in-situ heating*

## *Highlights*

1. Temperature-dependent bulk plasmon energy of tungsten can be used for temperature measurement from RT to 700 ºC in an *in-situ* heating experiment.
2. Variations in sample thickness of FIB-prepared electron-transparent TEM samples lead to uncertainties in temperature determination.
3. By exploiting the beam broadening of the bulk plasmon peak at lower sample thicknesses, the thickness effect can be corrected for accurate temperature measurements.

## *Introduction*

Scanning / Transmission electron microscopy (S/TEM) has been established as a powerful method for atomic-scale material characterization [1]. The further advancement of *in-situ* capabilities to expose materials to relevant process and application conditions opens access to structure-property correlations in materials [2], [3], [4]. Particularly, the development of microelectronic mechanical systems (MEMS) as *in-situ* sample stages [5], [6] enable real-time observation of material under appropriate stimuli such as heating [7], [8], electric biasing [9], [10], mechanical deformation [11] etc. in S/TEM experiments with a resolution up to the atomic scale [12], [13].

For example, S/TEM heating experiments have been widely applied for understanding thermally induced microstructure dynamics in materials such as phase transformations in metals (e.g. [14], [15]). To accurately correlate the findings from *in-situ* heating experiment to other microstructure information, e.g. to correlate results with predictions from phase-temperature or time-temperature diagrams, it is important to measure the temperature *in-situ*, accurately across the TEM sample. Thus, it is imperative to develop methods for *in-situ* temperature measurement [16], [17]. In particular, this allows for precise control of a temperature profile during an *in-situ* heating experiment to introduce specific thermal conditions, e.g. introducing temporal and spatial thermal gradients [18] that mimic the non-equilibrium thermal conditions e.g. in additive manufacturing processes. Therefore, accurate local temperature measurements across the sample are required to achieve controllability of temperature settings in *in-situ* heating experiments with nanometer resolution across micrometer-sized TEM samples.

Over the years, various approaches have been developed to measure the temperature of a TEM sample [16], [17]. For example, several thermometric methods utilize properties of thermometric materials, such as temperature-dependent sublimation rate of silver (Ag) particles (with a temperature accuracy of ± 5 °C) [19] or a phase transition of vanadium dioxide ($VO_2$) nanowires [20]. These characteristic temperature points can provide a direct binary temperature measurement; however, the application is limited by size dependencies, nano-size effects of phase transition and potential effects of the local environment, etc. Further, electron diffraction (ED) techniques have been applied to determine the temperature based on the thermal expansion of the material [21], [22] allowing e.g. different measurement points across the reference sample of a thin Au film (with a temperature accuracy of ± 2.8 °C) [22]. The measurements of temperature-dependent changes in ED patterns require a high phase stability in materials and a critical alignment of the TEM to maintain a highly parallel beam illumination [22]. Another method to measure the temperature uses the frequency shift of Raman peaks in materials in correlation with the sample temperature, however this requires a complex integration of a Raman spectrometer to measure the temperature on a sample inside a TEM [23]. By means of the principle of detailed balance sample temperature can also be measured using STEM-EELS. This principle states that the ratio between the probability of an electron exciting the sample from its initial energy state ($E_1$) to a higher state ($E_2$) and the probability of the sample de-exciting in opposite direction, giving energy to the probing electron, is given by $\exp(-(E_2 - E_1)/(k_B T))$. This temperature dependent ratio is retrievable from an EELS spectrum by measuring the ratio between a corresponding couple of loss and gain peaks. Lagos et al. [24] measured the temperature in a MgO nanocube this way with an uncertainty of ± 1 °C using phonon loss and gain peaks. In a similar way Idrobo et al. did measurements on h-BN nanoflakes [25]. Although this method allows for a direct (calibration-free) way of probing for local temperature it requires a highly monochromated beam with an energy resolution in the meV range. Currently only a handful of instruments can provide this making it inaccessible for most TEM users. Other studies [26] have used confocal microscopy and exploited luminescence spectra of nanoparticles for temperature measurement with a temperature measurement accuracy of better than 4 °C. However, the application temperature is limited ranging only up to 227 °C. Similarly, the cathodoluminescence frequency in semiconductors has been applied for temperature measurement. Here the drawback of the approach is a high sensitivity to heterogenicities in the local composition, causing an uncertainty of up to 50 °C [27].

Here in our study, we decided to investigate the accuracy of plasmon energy expansion thermometry (PEET) [28], [29], [30], [31], [32], [33], [34], [35], [36], [37], [38], [39], [40] as the method to measure local temperature in a S/TEM sample using STEM-EELS. PEET has been introduced in the past [37] as a non-contact thermometric method capable of mapping local temperature profiles across a S/TEM sample with nanometer spatial resolution [37].

A bulk plasmon describes the collective oscillation of the valence electrons within the material excited by an external electric field. The main principle of PEET is therefore to exploit the temperature dependence of a material's bulk plasmon energy *(Ep)*, expressed by:

(1) $E_p = \hbar \sqrt{\frac{ne^2}{m\varepsilon_0}}$

according to the free-electron model, where $\hbar$ is the reduced Planck constant, $\varepsilon_0$ the permittivity of vacuum and *n* is the density of valence electrons with charge *e* and mass *m*. The temperature dependence of the plasmon energy arises from the principle of thermal expansion, which affects *n* [41]. Consequently, a prerequisite for PEET is that in the temperature range of interest, no phase transformation nor abrupt morphological or structural change may occur inside the sample, as this would lead to a sudden irreversible alteration of *n*.

Previous work on PEET has demonstrated the temperature-dependent shift in *Ep* in various materials such as polycrystalline tin (Sn) film of 2.5 mm diameter [42], silicon (Si) nanoparticles [43], aluminium (Al) thin films [44] and silver (Ag) films [38] with increasing temperature. In more recent years, the focus in method development has been on pushing the spatial and temperature resolution achievable with PEET. In 2015, Mecklenburg et al. [37] mapped an inhomogeneous temperature distribution ranging from RT to 327 °C (with a thermal gradient of the order of $10^4$ °C/m) across a 80 nm-thick serpentine Al wire with a length of 0.8 µm, achieving a spatial resolution of 3 nm limited by the delocalization length of Al plasmon. In a subsequent study in 2018 [36], they incorporated drop-casted Si nanoparticles of 90 nm in size to expand the applicability of PEET from RT to 1250 °C. However, the low thermal expansion of Si (zeroth order thermal expansion coefficient of 3.3 × $10^{-6}$ °C$^{-1}$) led to low sensitivity to measure a temperature change. Similarly, Chmielewski et al. [35] used dry-deposited 50-nm-sized Al nanoparticles to measure the local temperature on a MEMS heating chip in vacuum and in hydrogen (H$_2$) environment. Shen et al. [30] studied molybdenum disulfide (MoS$_2$) with a large thermal expansion coefficient of 1.9 × $10^{-5}$ °C$^{-1}$ to measure thermal gradients of 8 × 10$^7$ °C/m (temperature difference of 200 °C over 2.1 µm length). They reported a temperature precision of 20 °C achieved by averaging measured plasmon energies along a column of pixels transverse to the gradient. This temperature precision in combination with the known thermal gradient resulted in a spatial resolution of 250 nm. Barker et al. [28] reported on measurements on Si nanoparticles using an optimized and automated plasmon peak fitting scheme allowing to live-process the average temperature over a selected region of interest (ROI). By collecting sufficient statistics from a homogeneously heated ROI, they have reported sub-50°C resolution. Lastly, just recently Kumar et al. [32] have exploited PEET to measure cryogenic temperatures in cryogenic TEM holders using again Al as specimen material.

So far, PEET has been applied to measure temperature in (clusters of) nanoparticles, thin films and various two-dimensional (2D) materials, but a potential influence from sample thickness on plasmon energy ($E_p$) has not been systematically investigated yet. To the best of our knowledge, there are only a few details in previous studies addressing a possible sample thickness effect on $E_p$. Mecklenburg et al. [37] reported that the local thickness variation from 0.03 to 0.34 relative thickness ($t/\lambda$) (within an otherwise homogeneous 80 nm-thick Al wire) does not show a correlation with $E_p$. Mitome et al. on the other hand [43] states an inverse square relation of $E_p$ to Si nanoparticle cluster sizes ranging from 3.5 nm to 10 nm. They conclude that this relation is caused by a quantum confinement effect. Similarly, Hu et al. [34] addressed the difference in thermal expansion within the different layers of 2D materials, such as graphene, $MoS_2$, $MoSe_2$, $WS_2$ and $WSe_2$, also because of quantum confinement effects. These studies suggest that $E_p$ might be affected by the sample thickness, especially in TEM lamellas with typical thicknesses well below 100 nm, e.g., when prepared by focused ion beam (FIB) [10], [18], [45], [46], with potential deviation in thermal expansion compared to bulk values caused by surface effects or induced-strain effect. Therefore, sample thickness is believed to have an impact on the local $E_p$, especially when mapped across a typical TEM lamella with varying thickness.

The aim of this paper is to investigate further this possible dependence of plasmon energy on sample thickness as this would have a significant effect on the accuracy of PEET, especially when applied to specimens with inhomogeneous sample thickness. Tungsten (W) was selected as the model material, because it is stable and is the metal with the highest melting point (at atmospheric pressure bulk tungsten melts at 3410°C [47]). This choice delays temperature induced morphological changes and melting effects in the heating experiments. In addition, W is known to exhibit a sharp plasmon resonance [48], [49], [50] which leads to high signal-to-noise ratios (SNR) in plasmon energy measurements compared to other materials and enables accurate $E_p$ determination. A W lamella with varying thickness was prepared and STEM-EELS mapping has been exploited to measure local sample thickness ($t/\lambda$ - log-ratio method) as well as corresponding local bulk plasmon energy at different temperatures. The experimental results showed a clear correlation between specimen thickness and plasmon energy at all set temperatures. It was found that the plasmon energy deviates more from theoretical expected values at the thinner region. It is shown that PEET can however still be performed by using thickness-dependent thermal expansion coefficients. Finally, an alternative calibration method is proposed, where *Ep* and the simultaneously observed broadening of a plasmon peak are combined into a thickness-independent parameter that is, however, still temperature-dependent to retrieve the local temperature across a TEM sample with varying thickness.

## *Method*

### *TEM sample preparation for in-situ heating experiment*

A Helios 5 Hydra UX plasma focused ion beam (PFIB) DualBeam® system (Thermo Fisher Scientific Inc.) was used for the TEM sample preparation. The W lamella used for this study was lifted out from a disposed W Easylift® needle. A xenon (Xe) plasma beam at 30 kV and beam current 16 nA to 4 nA was used in the first step of trenching, while for cross section thinning 30 kV and 1 nA were used. To polish the sample until electron transparency (< 100 nm), Xe plasma beam setting of 5 kV and 30 pA were used, and the sample was tilted from 3 to 7 degree to achieve a thickness gradient at the centre of the W lamella. After polishing, to minimize the thickness of the amorphous surface layer, the W lamella was lifted out using the Easylift® needle with W gas injection at plasma beam of 30 kV and 0.1 nA for sample attachment. The lamella (with dimension of 5 μm by 13 μm) was positioned over a through-hole window of a Wildfire® heating chip (DENS Solutions B.V.). This chip was placed on a 45° pre-tilt stab [45]. For process examination during the lift-out process, secondary electron (SE) imaging in scanning electron microscopy (SEM) mode using the electron beam was performed at 30 kV and 30 pA.

The through-hole type Wildfire® heating chip allows heating to temperatures ranging from RT up to 1300 °C with a stated accuracy of 5% at the centre windows [51]. This is achieved by Joule heating of a metallic heating spiral connected to a 4-point-probe sensing setup that monitors its temperature precisely by measuring its temperature-dependent resistance. A low sample drift rate down to 0.1 nm/min enables high stability and high spatial resolution during heating experiments [51]. The chip was used in combination with a dedicated Wildfire single-tilt® TEM heating holder.

### *EELS spectra acquisition in STEM mode*

The STEM-EELS acquisitions were performed using a monochromated Titan X-FEG® (Thermo Fisher Scientific Inc.) operated at an accelerating voltage of 300 kV. The EELS maps in STEM mode were collected using a Quantum 966® Gatan Image Filter (GIF) with DualEELS® capability and a US1000XP® camera. The Wien-filter monochromator allowed an energy resolution of 130 meV ± 10 meV, as measured by the full-width at half maximum (FWHM) of the zero-loss peak (ZLP). All EELS measurements were performed in micro-probe mode at a beam current of approximately 0.1 nA using a semi-convergence angle $\alpha$ of 1.90 mrad and semi-collection angle $\beta$ of 4.60 mrad. The selection of these specific $\alpha$ and $\beta$ values was based on preliminary measurements using various combinations of $\alpha$ and $\beta$ (details of the exact determination of $\alpha$ and $\beta$ can be found in ***SI.1***), aiming for optimal measurement of *Ep* (see details in ***SI.2***).

### *Estimation of absolute sample thickness*

The EELS map for thickness estimation was taken at 20 °C of the rectangular ROI including the thickness gradient of interest. The thickness in units of inelastic mean free path ($t/\lambda$) can be obtained by EELS and the log-ratio method [50], [52] with formula $t/\lambda = \ln(I_t/I_0)$. Here $I_t$ is the total transmitted intensity (including ZLP) and $I_0$ is the intensity corresponding to ZLP. In order to estimate $t/\lambda$ in the W as accurate as possible, a relatively high energy dispersion of 0.25eV/pixel was used to cover a wide range of energy loss (i.e. from - 35 eV to 477 eV).

To determine the absolute thickness value, the inelastic mean free path in W has to be estimated as well. Here the formula from [52] was used in combination with an effective semi-collection angle ($\beta^*$) to take the incident-probe convergence into account:

$$(2) \quad \lambda = \frac{106 F E_0}{E_m \ln\left(\frac{2\beta^* E_0}{E_m}\right)}$$

where the result is expressed in units of nanometers and $F = [1 + (E_0/1022)]/[1 + (E_0/511)]^2$ is a relativistic factor equal to 0.513 for incident electron energy $E_0$ of 300 keV and $\beta^*$ is the effective collection semi-angle. $E_m$ is an average energy loss defined by $E_m \approx \int S(E)\, dE / \int S(E)/E\, dE$ where $S(E)$ is the single scattering distribution (the experimental spectrum with plural scattering removed). These formulas are derived from electron scattering theory and more specifically from the Kramers-Kronig sum rule [53], [54]. To estimate $E_m$, Malis et al. [52] determined a phenomenological relation

$$(3) \quad E_m = 7.6 Z^{0.36}$$

in function of atomic number Z. The absolute thickness was estimated using Digital Micrograph 3.20 (Gatan Inc., New York, USA) where $I_0$ was determined by extracting the ZLP using the reflected tail model [55]. The provided parameters were $E_0$ = 300 keV, $\alpha$ = 1.90 mrad, $\beta$ = 4.60 mrad and Z (for W) = 74.

*W bulk plasmon peak characterization in STEM-EELS*
The PEET measurements were conducted at a set temperature of 20 °C (RT), then 150 °C, 200 °C, 250 °C and continuing with increments of 50 °C up to 1000 °C. The same ROI for collecting the W bulk plasmon peak was chosen to include the thickness gradient, which also is included in the absolute thickness map, as described in the previous paragraph. The STEM-EELS maps were taken in DualEELS® mode at the dispersion of 0.01 eV/pixel. The first spectra contain the ZLP in the energy range from approximately - 7.00 eV to 13.47 eV. The second spectrum tracked down the W bulk plasmon peak from 13.00 eV to 33.47 eV. (These two spectra are referred to as the *ZLP spectrum* and the *plasmon spectrum*, respectively, in the remainder of the document.) Besides the advantage of collecting both the ZLP and the plasmon peak at the dispersion of 0.01 eV/pixel, DualEELS® also allows recording the plasmon spectrum at longer exposure times in order to increase SNR. The exposure times at each probe position were set to 0.002 s and 0.098 s, respectively, corresponding to a total dwell time of 0.1 s. For mapping a step size of 31 nm in x and y was chosen. For more details about ROI corrections between acquisitions see **SI.3**.

The acquired data was processed and analysed using the python pyEELSMODEL module [56]. First the energy axis on every pair of ZLP spectrum and plasmon spectrum were aligned to set the ZLP at 0 eV of energy loss. Afterwards both the ZLP and plasmon peak were fitted in order to determine their respective peak position with sub-pixel precision. The considered fitting windows ranged from - 0.5 eV to 0.5 eV for the ZLP and from 17.5 eV to 30 eV for W plasmon peak, respectively. The trust-region-reflective least squares fitting algorithm was used. Several fitting models suggested by literature [28], [35], [36] were investigated on both peaks and compared based on their reduced $\chi^2$-value maps (details are shown in the *SI.4.1*). Based on the reduced $\chi^2$-value maps, a Voigt curve fitting was used for fitting the ZLP and a Johnson's $S_U$ curve for fitting the plasmon peak. The Johnson's $S_U$ curve has five parameters and can be asymmetric, this resulted in good fits at all probe positions even in cases when the plasmon peak was asymmetric when probing thickness gradients (details in *SI.4.2*). Unlike in case of the Voigt curve fitting, none of the Johnson $S_U$ curve parameters however represent its mode directly, so this was determined by numerically finding the root of its (analytical) derivative. To determine the plasmon energy, the fitted mode of the plasmon peak was simply subtracted from the fitted centre of the ZLP. Finally, broadening of the plasmon peak was determined by determining its full-width half maximum (FWHM) calculating the distance between the roots of its fit subtracted by half its max value. The FWHM values has been used to monitor the broadening of the plasmon peak and thus its potential correlation with changes in sample thickness. All numerical root-finding for both determination of plasmon fit mode and FWHM were performed using 'fsolve' function from the scipy.optimize python module [57]. This function uses a modified Powell's dog leg method. The plasmon energy maps and FWHM maps at the different set temperatures are shown in *figure S7* and *figure S8* respectively (see *SI.4.3).*

*Determining $I_t$-map and relative thickness map*
In addition to the *Ep* and FWHM maps, two more maps were derived from the DualEELS® data, namely the total transmitted intensity ($I_t$) map and relative thickness maps. For both maps, the ZLP and plasmon spectrum at each probe position were merged to form one spectrum. However, since the exposure times and hence the received dose of these two spectra differed by a factor of ∼ 48 and the two spectra had an overlap range at 13.0 eV to 13.47 eV, a direct combination was not possible. The overlapped region of both spectra showed a flat signal suited to determine the exact ratio between the averages of the two spectra. The plasmon spectrum was divided by this ratio, and the overlapping range (13.0 eV to 13.47 eV) was cropped in the ZLP spectrum due to the higher shot noise compared to the shot noise in the plasmon spectrum. At last, the two spectra were combined, and the energy axis aligned to have the ZLP's mode at 0 eV. $I_t$ maps were then extracted allowing the monitoring of possible morphological changes with increasing temperature (*SI.5*).

The relative thickness map was calculated using $\ln(I_t/I_0)$ where $I_0$ was taken as the total counts of a Voigt-fitted ZLP within an energy interval from - 3.0 eV to 3.0 eV again using the trust-region-reflective least squares fitting algorithm. Although this formula is the one used for calculating the specimen thickness in units of inelastic mean free path ($t/\lambda$), here it is not giving accurate values as the energy range of the spectrum is limited to less than 33.0 eV (after alignment of ZLP). The value does however give accurate relative differences in thickness and can among others be used for monitoring integrity of the thickness profile of the lamella with increasing temperature.

Due to the symmetry of the specimen thickness along the vertical axis (y-axis) in the mapped ROI, the remainder of this paper primarily focuses on the averaged horizontal line profile of all the different attained maps (absolute thickness, *Ep*, FWHM as well as *t/λ* maps). By averaging over the 16 pixels along the y-axis at every probe position along the x-axis, statistical weight was gained, and a corresponding standard error on the considered signal was determinable. The obtained average relative thickness profiles from the $\ln(I_t/I_0)$ maps allowed to determine final corrections in horizontal shift to properly align the increasing thickness for all set temperatures. Slight misalignments of 124 nm at the maximum (4 pixels × 30 nm) originate from slight mismatch in mapped region between the acquisitions at different set temperatures. The horizontal shift corrections were applied to all maps, after which they were cropped to a width of 93 pixels map instead of the originally 100 pixels width. Minor vertical shifts were also noted between maps but since the thickness symmetry along y axis extended in regions above and below the ROI this was found to be negligible.

Finally, it was verified that the average relative thickness profiles attained from the DualEELS® measurement have a linear relation to the average absolute thickness profiles attained from the 0.25 eV/pixel dispersion in the EELS measurement. Therefore, a linear relation between the relative thickness values measured at 20°C and absolute thickness values (also measured at 20°C) was derived. This linear relation was then used to transform all relative thickness values at all other set temperatures to absolute thickness values. It must be noted that the obtained absolute thicknesses at temperatures higher than 20°C are only approximately correct as thermal expansion effects are not accounted for here.

*Simulation sample temperature using finite element method (COMSOL)*
To predict the potential temperature distribution in the W lamella deposited on the MEMS heating chip, finite element method (FEM) simulations were conducted using COMSOL Multiphysics® software [58]. These simulations, based on heat transfer theory in solid mechanisms, included three modules: electric current, solid mechanics, and heat transfer in solids. All relevant aspects of the physical system and interactions among the modules were simulated simultaneously, modelling from the Joule heating principle to heat transfer across the entire geometry. To mimic the vacuum environment in TEM, only radiative heat loss and heat flux from conduction were assumed in the heating simulation.

The finite element model consisted of the free-standing membrane area with the embedded heating spiral and the W lamella attached to it. The thickness of the lamella was also included into the model. A tetrahedral mesh was used with a mesh size from 0.2 um to 10.5 um, and mesh rate at the sample area was set to 0.2. The thermal properties of W were taken from the COMSOL library with the coefficient of thermal expansion (CTE) of $4.5 \times 10^{-6}$ °C$^{-1}$, thermal conductivity of 175 W/(m·K) and the emissivity of 0.361. The electric voltage was ramped from 0.746 V to 1.421 V in correlation with the set temperatures ranging from 300ºC to 1000ºC (shown in figure ***S8b***), with the set temperature defined by the temperature at the centre of the MEMS heating spiral. In order to evaluate the accuracy of our simulations, a separate model was made consisting of just the free-standing membrane area with the embedded heating spiral without lamella. Simulations of this model at different set temperatures showed good agreement with the values reported by van Omme et al. [51] (more details in ***SI.6)***.

## Results

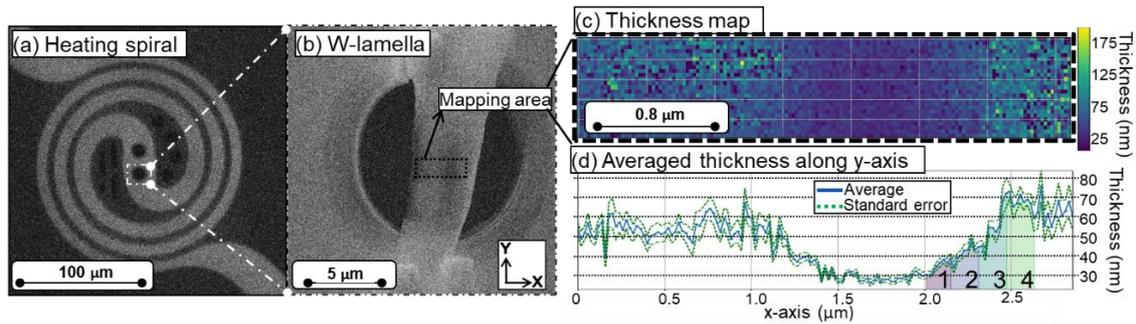

*Figure 1: Experimental setup of the TEM heating experiments and the thickness map of W-lamella. SEM SE images show (a) MEMS heating spiral and (b) W-lamella placed over the central window of the magnified area in (a). (c): The thickness map is based on log-ratio (t/λ) method, corresponded to the assigned area in (b). (d): Line profile shows the average thickness variation of the W lamella with standard error, along the y-axis in the thickness map of (c). Four different thickness regions are indicated in (d) by a colour code with numbers 1 to 4.*

Figure 1a shows a SEM image of the heating spiral of the MEMS heater using secondary electrons (SE), and the higher magnification SE image in figure 1b shows the W-lamella deposited over the through-hole window of the MEMS heater (as the dash-lined square in figure 1a). The ROI around the thickness gradient that got mapped during the heating experiments is indicated by a dash-lined rectangle in figure 1b. The absolute thickness (attained from the EELS measurements at 20 °C prior to heating, using 0.25 eV/pixel energy dispersion) for this ROI is shown in figure 1c. The line thickness profile shown in figure 1d indicates varying thickness between approximately 30 nm and 70 nm which is the range of thicknesses present in most FIB-prepared electron-transparent TEM samples [45], [59], [60], [61]. The error bars for the values are standard errors from averaging every column along the vertical axis transient to the thickness gradient.

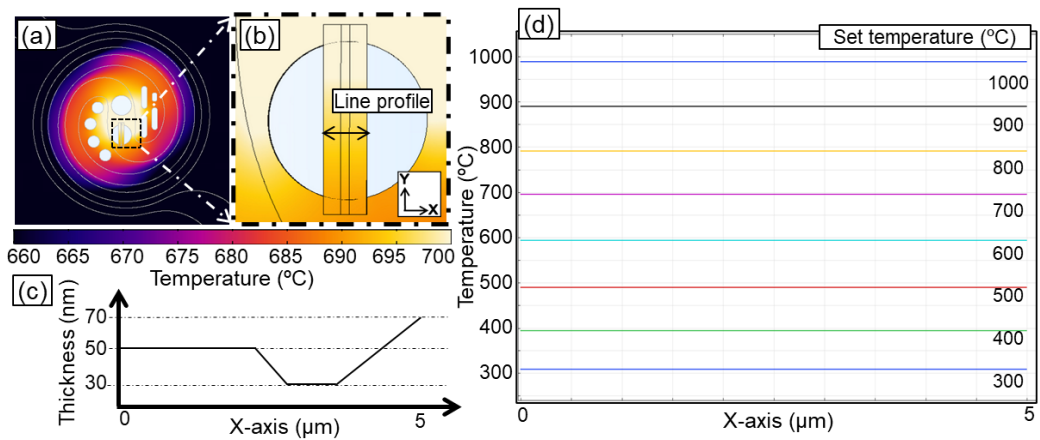

*Figure 2: FEM simulations showing temperature distribution in the W lamella with varying sample thickness. Simulated temperature distribution at set temperature of 700°C of (a) the entire geometry including free-standing membrane with embedded heating spiral and W lamella (b) the attached W-lamella over the magnified area in (a). (c): Schematic diagram of the considered line thickness profile of the sample geometry along the double-arrow line in (b). (d): Line temperature profiles along the double-arrow line in (b) at set temperatures from 300°C to 1000 °C with steps of 100 °C.*

The results from the FEM simulations (see details in *Methods*) are shown in figure 2. As an example, for a set temperature of 700 ºC at the centre of the spiral, the temperature distribution over the entire geometry is shown in figure 2a. Zooming in at the W lamella (figure 2b) the temperature reaches between 690 ºC and 700 ºC. Such deviations in temperature between setpoint and sample area of up to 5 % are within the accuracy of the vendor calibration [51] and have been reported before in actual heating experiments [18]. The considered thickness profile along the double-arrowed line in figure 2b is shown in figure 2c, resembling the measured thickness profile of the actual sample as depicted earlier in figure 1d. Temperature line profiles along the double-arrowed line show homogeneous temperature across the W lamella for all set temperatures from 300 to 1000 ºC (with step of 100 ºC). These FEM simulations indicate that temperature should be independent of the specimen thickness variations between 30 nm and 70 nm. As mentioned above, the deviation between the simulated temperature over the W lamella and the setpoints from 300 to 1000 ºC are all within the accuracy of 5 %.

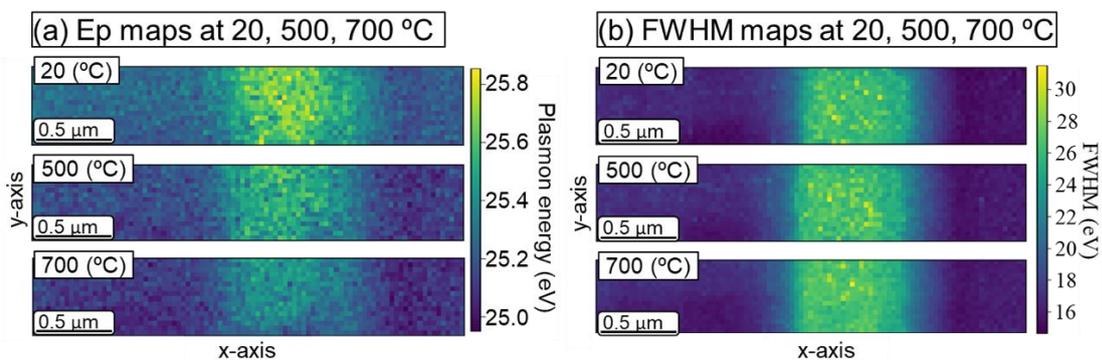

Figure 3: Maps of measured W Ep (a) and FWHM (b) over the W-lamella with varying thickness at set temperature of 20 °C, 500 °C and 700 °C. The mapping area is the same as the mapping area in figure 1c.

As explained in the *Methods* section, the DualEELS® measurements for the study of *Ep* were performed on the same ROI as in the absolute thickness map of figure 1c for temperatures ranging from 20 °C to 1000 °C. The resulting *Ep* and FWHM (= *Ep* peak broadening) maps are shown in respectively figures *S6* and *S7*. Due to noted morphological change at temperatures above 750 ºC (see *SI.5*), only data at set temperatures ranging from 20°C to 700ºC are included from now on (not limiting the main focus of our study), as morphological or structural changes lead to irreversible changes of *n*. and limit the applicability of PEET (see *Introduction*). This observed morphological change is most likely due to a potential recrystallization starting at this temperature in the thinnest part of the W lamella (for further discussion see the details in *SI.5*).

Maps of *Ep* and FWHM are shown in figure 3 for selected set temperatures of 20 °C, 500 °C and 700 °C, respectively. Figure 3a indicates two important trends. First, a red shift of *Ep* maps with increasing temperature, as expected from the principle of PEET. The average *Ep* value shifts from 25.35 eV to 25.18 eV with increasing temperature from 20 °C to 700 °C. Secondly, a thickness dependency is seen, shown by the *Ep* change (i.e. colour change to green/yellow in the *Ep* maps in figure 3a) around the thinnest part of the sample (i.e. at the approximated 30 nm sample thickness, see figure 1c). Even though a clear thickness dependence of the plasmon energy is seen, it can be noted that the measured average *Ep*, especially at 20°C, is in good agreement with earlier literature values of 25.3 eV [62], [63] and 25.5 eV [64], [65] reported for unspecified thickness and temperature condition.

Upon comparing to the corresponding thickness profile in figure 1d, it becomes more evident that *Ep* decreases with increasing thickness from the approximated 30 nm to 70 nm. At 20 ºC, *Ep* decreases by 0.43 eV from 25.63 eV to 25.20 eV. At 500 ºC, *Ep* decreases by 0.40 eV from 25.51 eV to 25.11 eV. At 700 ºC, *Ep* decreases more significantly by 0.31 eV from 25.38 eV to 25.07 eV. The FWHM maps in figure 3b show a thickness dependency at 20 ºC, 500 ºC and 700 ºC, with a larger FWHM value at the thinnest region. At 20 ºC, the FWHM increases from 15.25 eV (at 60-70 nm thickness) to 23.31 eV (at 30-40 nm thickness), at 500 ºC from 15.7 eV to 22.5 eV and at 700 ºC from 15.8 eV to 22.6 eV with decreasing the sample thickness.

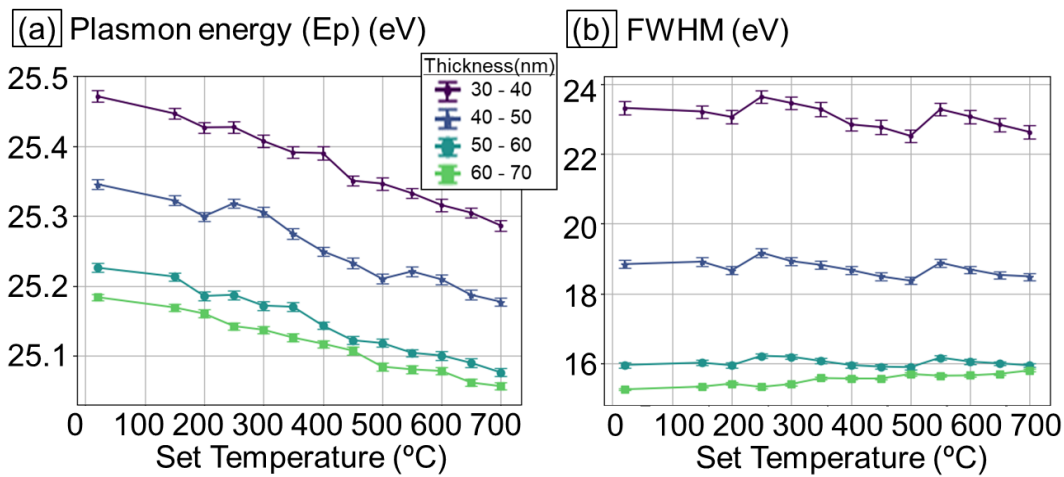

*Figure 4: (a) Average Ep and (b) FWHM of the plasmon peak in function of set temperatures from 20 °C to 700 ºC for different thickness ranges from 30 to 70 nm. The error bars correspond to standard error. Legend shared for (a) and (b) shows the corresponding thickness ranges, which are extracted from the thickness estimation, in figure 1d with assigned colour blocks and numbers.*

To quantitatively analyse the relationship of *Ep* and FWHM with sample thickness, values were extracted from their maps (figures 3a and 3b respectively) within four selected sample thickness intervals with different thicknesses shown at the thickness gradient area in figure 1d. The thickness intervals range approximately from 30-40 nm, 40-50 nm, 50-60 nm and 60-70 nm, indicated by four colours (and numbered 1- 4). Each dataset contains 80 points with 16 pixels in y-direction (width of 0.48 μm) and a 5-pixel interval along x-axis (length of 0.15 μm). Figure 4a shows the average values of *Ep* within each of these distinct thickness ranges, when ramping the set temperature from 20 ºC to 700 ºC. As mentioned before, it clearly shows a decrease in *Ep* with increasing sample temperature. Furthermore, it also indicates that the averaged plasmon energy *Ep* is higher in the thinner sample regions compared to thicker sample regions. For example, at 20 ºC, *Ep* decreases by 0.29 eV, from 25.47 eV (at 30-40 nm) to 25.18 eV (at 60-70 nm), respectively. At 500 ºC, *Ep* decreases by 0.27 eV, from 25.35 eV (at 30-40 nm) to 25.08 eV (at 60-70 nm), respectively. At 700 ºC, *Ep* decreases only by 0.23 eV, from 25.29 eV (at 30-40 nm) to 25.06 eV (at 60-70 nm). A more pronounced decrease in *Ep* is observed in the thinner sample areas with increasing temperature. From 20 ºC to 700 ºC, at thickness region 1 (30-40 nm) *Ep* decreases by 0.18 eV (25.47 eV to 25.29 eV), at thickness region 2 (40-50 nm) *Ep* decreases by 0.17 eV (25.35 eV to 25.18 eV), at thickness region 3 (50-60 nm) *Ep* decreases by 0.15 eV (25.23 eV to 25.08 eV) and at thickness region 4 (60-70 nm) *Ep* decreases by 0.12 eV(25.18 eV to 25.06 eV).

Figure 4b displays the average FWHM of W bulk plasmon peak at each thickness region at set temperature from 20 ºC to 700 ºC. It shows that the FWHM does not vary significantly with the change in temperature and increases with thickness at each fixed set temperature. In the thickness region 1 (30-40 nm), FWHM remains around 23.0 eV, at thickness region 2 (40-50 nm) FWHM remains around 18.5 eV, at thickness region 3 (50-60 nm) FWHM remains at 16.0 eV and at thickness region 4 (60-70 nm) FWHM remains at 15.5 eV. The shown error bars in figure 4a and 4b are calculated standard errors under the assumption that the thickness is constant within the intervals. Since this is actually not the case and the thickness in the intervals instead changes rather linearly along the x-axis, these assumed error bars are expected to be upper limit.

## *Discussion*

The results show a clear thickness dependence of *Ep* of W at the different set temperatures. This dependence affects the accuracy of the conventional PEET method which does not take thickness variations into account. In the following, this influence will be evaluated in more detail and approaches to take it into account are proposed.

1. *Temperature-dependent Ep in W - theory and experiments*
   - *Expected temperature-dependence of Ep in W - based on theory*

As mentioned in the *Introduction*, the common procedure of PEET is based on the free-electron model (eq.1) to derive temperature from *Ep* trough the temperature-dependant valence electron density *n(T)*. *n(T)* can be expressed in terms of coefficient of thermal expansion (CTE, $\alpha_l$) as:

(4) $n(T) = n(T_0) \cdot [1 - 3(\int_{T_0}^{T} \alpha_l(T) \cdot dT)]$

where *n(T₀)* is the density of valence electrons at an initial temperature $T_0$ and *n(T)* is the density of valence electrons at temperature *T*. Combining the eqs.1 and 4, the temperature-dependent plasmon energy can be expressed as:

(5) $E_p(T) = E_p(T_0) \cdot \left[1 - \frac{3}{2}\left(\int_{T_0}^{T} \alpha_l(T) \cdot dT\right)\right] = E_p(T_0) \cdot \left[1 - \frac{3}{2} \cdot \alpha_0 \cdot \Delta T\right]$

where $E_p(T)$ is *Ep* at temperature *T*. The right term can be to approximate a linear dependence of CTE ($\alpha_l$) into the zero order CTE ($\alpha_0$). Accordingly, the temperature-dependent shift in *Ep* can be calculated using values of the initial plasmon energy ($E_p(T_0)$) and the zeroth order CTE ($\alpha_0$):

(6) $\frac{\Delta E_p}{\Delta T} = -\frac{3}{2} \cdot E_p(T_0) \cdot \alpha_0$

where $\Delta E_p$ is $E_p(T) - E_p(T_0)$ and $\Delta T$ is $T - T_0$.

For comparison: in case of aluminium (Al), used by Mecklenburg et al. [37] with a bulk plasmon energy at RT of *Ep (T₀)* = 15.8 eV and a zero order CTE of $\alpha_0$ = 23.5 (10⁻⁶ K⁻¹), the shift in *Ep* for a 1 ºC change in temperature *ΔEp/ΔT* can roughly be estimated of being approximately 0.557 meV/ºC. For W considered here with the coefficient $\alpha_0$ of 4.31 (10$^{-6}$ K$^{-1}$) [66] and the average *Ep* of =25.35 eV at 20 ºC (figure 3a), the *Ep* shift per 1 ºC is expected to be around 0.164 meV/ºC. As a reminder: W was chosen because of the much higher melting temperature compared to Al to expand the applicability of PEET to higher set temperatures. (*Note: The estimation of the CTE value depends on the used measurements to determine the lattice constant [67]. Accordingly, the error bar is assumed to be less than 0.001 Å for a lattice constant of W of 3.164 Å, and, therefore, the error for CTE can be neglected.*)

- *Temperature-dependent Ep in W – based on our results*

If the measured thickness variations (see figure 1) are neglected and the average *Ep values* from the maps shown in figure 3a are used to estimate the average temperature dependence of *Ep*, then the average *Ep* decreases from 25.35 eV at 20 ºC to 25.18 eV at 700 ºC. This corresponds to an energy shift with temperature of ~ 0.25 meV/ºC, being 1.5 times larger than the above-estimated value of 0.164 meV/ºC. This already illustrates how large the influence of thickness variations can be on the accuracy of PEET if they are not taken into account.

- *Temperature-dependence of W Ep with thickness consideration*

The relation of the induced systematic error on the measured temperature, using the free-electron model with bulk literature value for $\alpha_0$ [66], and the sample thickness was further investigated by again considering the four thickness intervals indicated in figure 1d.

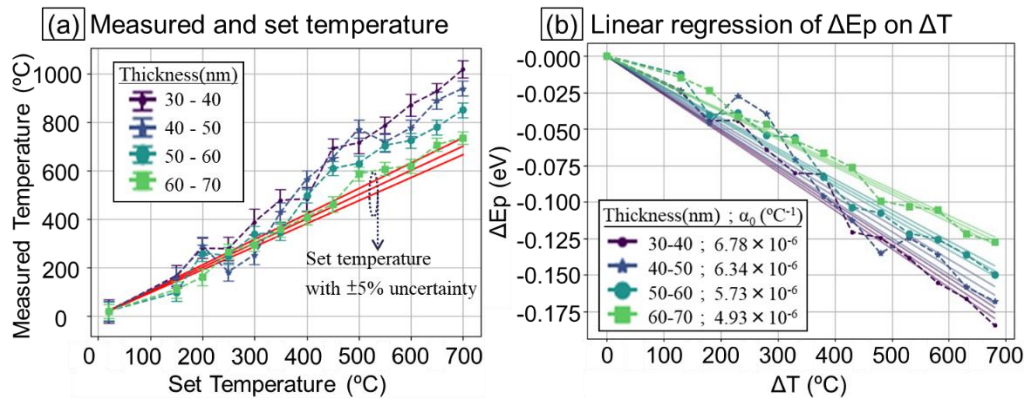

*Figure 5: (a): Measured temperature vs set temperature, according to the free-electron model using the literature bulk W $\alpha_0$ value, for the different thickness regions highlighted in figure 1d. (b): Linear regression of $\Delta E_p$ on $\Delta T$ and extracted $\alpha_0$ values for the different thickness regions considered. For every thickness interval three straight lines are represent. The central line is the linear regression line and the other two represent upper and lower limit.*

The results are shown in figure 5a, where the measured temperature is plotted against the set temperature. The measured temperature at the thick regions (60-70 nm) is in good agreement with the set temperature within the expected 5 % uncertainty [51]. With decreasing sample thickness, the deviation between measured and set temperature increases. At the sample thickness of 30-40 nm, the deviations are approximately 214 ºC and 317 ºC at set temperatures of the heating chip of 500 ºC and 700 ºC, respectively. The larger deviation in the thinner area implies the thermal expansion varies with sample thickness. The zeroth order coefficients ($\alpha_0$) were extracted for the different thickness regions using linear regression of $\Delta E_p$ on $\Delta T$, as shown in figure 5b. In figure 5b, the slope is more negative at a smaller sample thickness, i.e. from 0.259 meV/°C in the thickness region of 30-40 nm compared to 0.186 meV/°C in the thickness region of 60-70 nm, which is the closest to the theoretical value of 0.164 meV/°C (see section before). Accordingly, the deduced $\alpha_0$ has a larger value in the thinner area, from 6.78 ×10$^6$ °C$^{-1}$ at the thickness of 30 to 40 nm, 6.34 × 10$^6$ °C$^{-1}$(40-50 nm), 5.73 × 10$^6$ °C$^{-1}$(50-60 nm) to 4.93 × 10$^6$ °C$^{-1}$(60-70 nm), respectively. Similarly, compared to the literature value of $\alpha_0$ of 4.31×10$^{-6}$ °C$^{-1}$[66], the deviation becomes larger with decreasing sample thickness. This finding agrees with previous studies [68], [69], [70] showing that thermal expansion decreases with increasing size of the nanostructure. Pathak et al. [68] developed a theory for prediction and showed the changes in thermal expansion depends on the bounding surface of the nanostructure. The other study claimed that the change in thermal expansion of the thin films is attributed from the residual stress imposed from the mismatch of the film and the substrates[69], [70].

In addition to comparing the temperature dependence of *Ep* and extracting $\alpha_0$ in different thickness ranges, the temperature resolution was also calculated by propagating the error of $\alpha_0$ and $E_p(T_0)$ using eq.4 to present the measured temperature resolution achieved. The standard error found in the thickest region (60 to 70 nm) is lower than 41°C and increases slightly with decreasing thickness. For thickness region of 50-60 nm the upper limit is 43°C, 48 °C for the 40-50 nm thickness region and 49 °C for the 30-40 nm thickness region, respectively.

2. *Estimation of the effect of sample thickness on W bulk plasmon peak Ep and its FWHM*

To further investigate the trend of the effects of sample thickness on the W bulk plasmon peak in EELS, we sorted the values of *Ep* and the peak's FWHM in bins of absolute thickness. For every bin, the values were averaged, and the corresponding standard error were determined. The plots in figure 6 show *Ep* vs absolute thickness and FWHM vs absolute thickness. The absolute thicknesses are approximated values and were retrieved by calibrating the relative thickness maps as explained in the *Method* section.

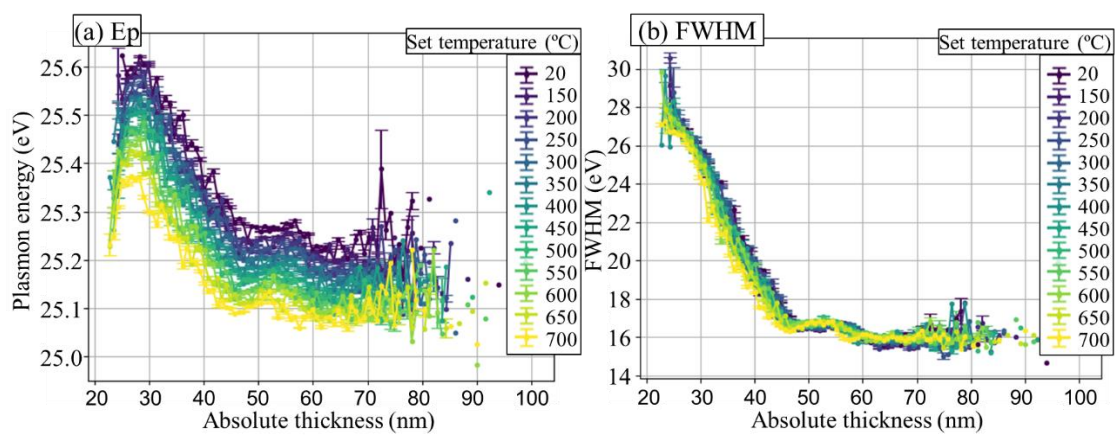

*Figure 6: approximated absolute thickness vs (a) Ep and (b) FWHM at set temperatures ranging from 20 °C to 700 °C as shown in the legend.*

- *Effect of sample thickness on bulk plasmon energy (Ep)*

Figure 6a shows that the *Ep* first increases from 25.58 eV at 20 °C and 25.23 eV at 700 °C to 25.62eV at 20 °C and 25.37 eV at 700 °C in the region below approximately 30 nm (energy value with an error bar of 0.01 eV). Above a thickness of approximately 30 nm, *Ep* starts drastically decreasing from 25.62 eV to 25.31 eV at 20 ºC and from 25.37 eV to 25.10 eV at 700 ºC with increasing sample thickness up to a thickness of 40 nm. In the sample thickness range from above 40 nm to approximately 60 nm, the decrease in *Ep* is less drastic with increasing thickness. Overall, the decreasing part can be approximated as decreasing by ~1/(absolute thickness). As outlined before in figures 3 and 4, the *Ep* is seen to increase with decreasing sample thickness and decreases with increasing set temperature.

Previous reports indicate that interband transition can potentially also contribute to the shift in plasmon energy [64], [71]. However, in our study, there is no pronounced peak present in EELS in the energy region of 20.5 eV to 28.5 eV that would indicate any presence of interband transitions.

From our viewpoint, one of the reasons for our observed deviation may be strain in the TEM lamellas caused by the lamella deposition on the MEMS heater. This argument is supported by the simulated stress distribution shown in *SI.7*. Previous work [72] has indicated that the induced strain in a thin film can be extracted from the bulk plasmon energy. Therefore, we assume that the sample thickness variation of the FIB lamella could cause an inhomogeneous stress distribution, especially introducing a higher stress in the thinner sample area, causing a difference in lattice expansion locally and a further deviation in the values of *Ep*. To quantify such a strain effect requires further studies that are outside of the scope of this work. The inverse behavior of *Ep* below a sample thickness of approximately 30 nm still needs further study to track the change of *Ep* in this thickness range. It is known that these datapoints originate from corner regions in the thinnest part of the lamella, i.e. close to x=1.6 *μm* and 2.0 *μm* in figure 1d.

- *Effect of sample thickness on plasmon peak broadening (FWHM)*

In addition to the value in *Ep*, the broadening (FWHM) of W bulk plasmon EELS peak also varies with local thickness, as shown in figure 3b, and indicates a temperature independence, as shown in figures 4b and 6b. In the thickness range from approximately 30 nm to 60 nm, the average plasmon FWHM decreases from around 26.0 ± 1.0 eV to 16.0 ± 1.0 eV and further saturates at 16.0 ± 1.0 eV at higher sample thicknesses. An approximate decrease of ~1/(absolute thickness) is observed, just as for *Ep* in figure 6a. Moreover, the FWHM curve corresponds very well to the *Ep* curve, except for the onset at the smallest sample thickness (from 20-30 nm). The observed *Ep* peak broadening shows an asymmetric enhancement on the lower-energy side of the bulk plasmon peak, as shown in the top 4 spectra in figure *S5*, compared to the spectra in thicker regions in the bottom 4 spectra in figure *S5*. This asymmetric broadening is the reason for the steady increase of the FWHM with decreasing sample thickness. We believe that this can be attributed to the existence of a W surface plasmon positioned at 21 eV [65]. Noteworthy is the proximity of the W surface plasmon peak at 21 eV and the bulk plasmon peak at 25 eV in our low-loss EELS signal. Hence, this finding suggests that, as the sample's thickness decreases, the influence of the surface plasmon could become increasingly pronounced. However, despite the proximity of the surface plasmon peak at 21 eV, there is no large deviation in the *Ep* value determined by curve fitting for two reasons. One point is that with peak broadening, the bulk plasmon still remains dominant in the spectra, as indicated in figure *S5*, causing only neglectable changes in peak position. The other reason is that the Johnson's Su equation includes a potential asymmetry in the plasmon peak, and this reduces the effect of any deviation from a symmetrical peak shape. As shown in figure *S4a*, the high fitting quality with low reduced $\chi 2$ values does not vary with local thickness variation (i.e. with variations in FWHM).

*3. Delocalization length dominate the spatial resolution*

To minimize beam effects in the STEM-EELS experiments by minimizing STEM spot overlap during scanning, a larger pixel size of 30 nm/pixel was chosen for acquiring all EELS maps. Furthermore, we calculated the electron beam-induced heating to be less than 0.02 °C (as detailed in *SI.8*), a value well below the temperature resolution and thus negligible. With an estimated sample drift of less than 0.1 nm/min [51] and a collection time for a map of 160 seconds, the maximum spot drift can be estimated as less than 0.3 nm, thus can be neglected here as well. To further evaluate the spatial resolution in PEET, we compared pixel size of 30 nm to the operated beam size and the localization of the EELS signal. First, the spatial resolution in Cs-corrected and mono-chromated STEM-EELS is defined by the probe size [73]. Since we applied the micro-probe STEM mode for EELS data collection with a low α of 1.9 mrad, the diffraction-limited probe size is ∼ 1.26 nm. On the other hand, the localization of the EELS signal can be expressed approximately by [50]:

$$(5) \quad L = 0.52 \cdot \lambda/\theta_E^{3/4} = 2.4 \ nm$$

where $\lambda$ = 1.96 pm is the electron wavelength at 300 kV and $\theta_E$ = 0.04 mrad is the characteristic scattering angle for W plasmon peak at around 25.35 eV. When comparing all the length values, the pixel size of 30 nm for collecting EELS maps emerges as the primary limitation in our experiments[37], and fundamentally nanometer resolution should be possible for PEET technique.

*4. Criteria for an optimal sample thickness to perform PEET on tungsten*

Based on the principle of PEET, bulk plasmon energy shift with temperature origins from the thermal expansion in 'bulk' materials. The point has been verified by the indications in our experiments: the *Ep* value 'saturates' when the sample thickness is larger than approximately 60 nm in figure 6a and there is a match of the measured sample temperature with the set temperature of the heating chip at sample thickness range of 60-70 nm in figure 5b. This implies that when applying PEET, the particular material, i.e. TEM sample, has a minimum threshold sample thickness. As an indication from our results, when using PEET to measure the temperature in W, the sample should have a homogeneous thickness distribution with a thickness larger than a determined minimum threshold value (e.g. 60 nm for W), or the sample thickness effect should be corrected.

Here an attempt is made to remove the thickness dependency of *Ep* and introduces a parameter *ξ* that depends only on temperature. This was achieved with input of the *Ep* that is temperature- and thickness-dependent and the FWHM that is only thickness-dependent, as shown in figure 7. This new parameter *ξ* is defined by:

$$(6) \quad \xi = \frac{E_p}{\frac{FWHM}{24} + 25.35 \ eV}$$

As it can be seen in figure 7, the parameter $\xi$ is thickness independent except for the very low sample thicknesses below 30 nm. The deviation at thicknesses below 30 nm could be explained by the poor correlation between the *Ep* and FWHM curve shapes at this thickness region, as shown in figure 6. Further investigations on the parameter $\xi$, including a more detailed analysis of the FWHM-dependent correction of *Ep*, require further studies.

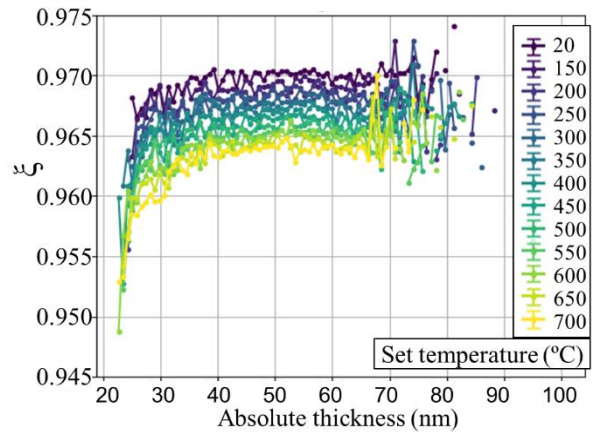

*Figure 7: The empirical value $\xi$ from Ep and FWHM with thickness. Below the approximated 30 nm, $\xi$ remains deviating with thickness dependence. Above this 30 nm, the $\xi$ shows only temperature dependence as aimed for.*

To optimize the application of PEET for temperature measurements of W lamella, these considerations have to be taken into account, as TEM lamellae typically have thicknesses in the range of a few 10s of nm. This most likely also applies for other materials. Accordingly, observed broadenings of the bulk plasmon peak will indicate increasing influences of the TEM sample surfaces, and the saturation region has the same thickness range as *Ep*, shown in figure 6. The deviation in values of *Ep* at thinner sample area could be 'mis-interpreted' as an indication of a higher local sample temperature, as indicated in figure 5a.

## *Conclusion*

Plasmon energy expansion thermometry (PEET) has been applied to measure local temperature variations in a PFIB-prepared W TEM lamella (with a sample thickness of < 100nm) that exhibits typical TEM sample thickness variations, including a thickness gradient from approximately 30 nm to 70 nm. This W lamella has been used to explore the effect of such a thickness non-uniformity on that bulk plasmon peak energy measured in EELS, as the temperature-depended measure in PEET, and a corresponding peak broadening (FWHM). The accuracy and uncertainty in temperature measurements using PEET has been further evaluated in *in-situ* S/TEM heating experiment with chosen set temperatures between 20 ºC (RT) and 700 ºC.

Based on our findings and discussions, the following conclusions can be drawn in case of applying PEET on a W TEM lamella:

- For a measurement of a local sample temperature using conventional PEET, the TEM sample thickness has to be larger than approximately 60 nm in case of W.
- The achievable temperature resolution can be in the range of ± 30 ºC.

- The value of the bulk plasmon energy *Ep* and peak broadening (FWHM) of the W plasmon peak are highly thickness dependent at sample thicknesses below 60 nm due to the increasing influence of surface effects and the induced stress concentration typically observed in such thin free-standing TEM lamellae.
- At lower sample thicknesses, the measured peak broadening of the W bulk plasmon peak has been used to correct for such thickness effects by introducing a parameter $\xi$.

## *Acknowledgement*

All authors acknowledge continued support by their colleagues at DTU Nanolab, EMAT, and NANOlab Center of Excellence. Y.-C. Y. would especially thank Murat Nulati Yesibolati for his support of the COMSOL simulations and Shima Kadkhodazadeh for fruitful discussions about the application of STEM-EELS. L.S. thanks Daen Jannis for his help with the use of pyEELSMODEL for the data processing. J.J. thanks DTU Nanolab for financial support (start-up grant). J.V. and L.S. acknowledge the eBEAM project which is supported by the European Union's Horizon 2020 research and innovation programme FETPROACT-EIC-07-2020: emerging paradigms and communities.

## *Declaration of competing interest*

The authors declare that they have no known competing financial interests or personal relationships that could have appeared to influence the work reported in this paper.

## *Author Contributions*

J. J. and Y.-C. Y conceived the initial project idea. Y.-C. Y. prepared the sample and N. G. acquired the STEM-EELS data. L. S. and Y.-C. Y. performed data analysis with help from J. J., J. V. and N. G. All authors were involved in discussions on the interpretation of the results. The manuscript was written with contributions from all authors.

## *Data availability*

The data that support the findings of this study are available from the corresponding author upon reasonable request.

**Supporting Information for: Improving the Accuracy of Temperature Measurement on TEM sample using Plasmon Energy Expansion Thermometry (PEET): Addressing Sample Thickness Effects**

Yi-Chieh Yang[1,&], Luca Serafini[2,3,&], Nicolas Gauquelin[2,3], Johan Verbeeck[2,3], Joerg R. Jinschek[1,*]

[1] National Centre for Nano Fabrication and Characterization (DTU Nanolab), Technical University of Denmark (DTU), Kgs. Lyngby, Denmark.

[2] Electron Microscopy for Materials Science (EMAT), University of Antwerp, Antwerp, Belgium.

[3] NANOlab Center of Excellence, University of Antwerp, Antwerp, Belgium.

[&] These authors contributed equally
[*] Corresponding author: jojin@dtu.dk

*SI.1 Calculation of the STEM convergence angle α and the EELS collection angle β:*

After the STEM alignment, prior to any heating experiment, the (semi-)convergence angle $\alpha$ was determined by acquiring a convergent electron diffraction (CBED) image on a highly crystalline part of the tungsten (W) lamella. The GIF entrance aperture was retracted (maximum collection angle) in order to prevent obstruction of first order diffraction spots. Image S1 shows the attained diffraction pattern taken along the [102] zone-axis. From the first order diffraction spots a value for $\alpha$ was calculated using:

$$\alpha = \arctan\left(\frac{R \tan(2\theta_{Bragg})}{2\Delta x}\right) = \arctan\left(\frac{R \tan\left(2 \cdot \arcsin\left(\frac{n\lambda}{2d}\right)\right)}{2\Delta x}\right)$$

Here $d$ is the corresponding interplanar distance of the considered spot determined by $d_{hkl} = a/\left(\sqrt{h^2 + k^2 + l^2}\right)$, given the body-centred cubic (bcc) structure of W, with lattice parameter of $a$=316.52 pm at 20°C [1]. $R/\Delta x$ is the ratio of the diameter of central spot ($R$) and the distance from central spot to considered first order spot ($\Delta x$). Both $R$ and $\Delta x$ were measured in units of pixels on the diffraction image. Further, n=1 and $\lambda$=1.96 pm (relativistic electron wavelength at 300 keV). The average $\alpha$ at last was 1.90 mrad.

Next, the GIF entrance aperture of 2.5 mm diameter was reinserted blocking the first order spots and casting a black shadow around the central spot. The (semi-)collection angle $\beta$ was then calculated by:

$$\beta = \arctan\left(\frac{D}{R} \tan(\alpha)\right)$$

Where $D$ is the diameter of the opening of the entrance aperture seen on the diffraction image measured in units of pixels just like R.

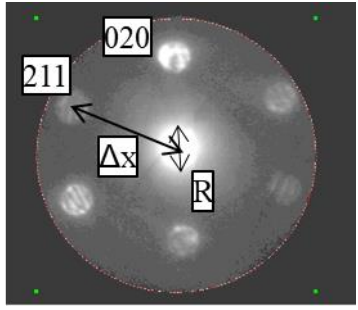

*Figure S1: CBED pattern taken along [102] zone-axis of W with GIF collection aperture retracted in order to determine α.*

### SI.2 Ep maps at room and at high temperature using different α and β:

To investigate the influence of different α (semi-convergence angle of the STEM probe) and β (semi-collection angle for EELS) on the measurement of the bulk plasmon energy *Ep* in EELS, we have used two different values for α (1.9 mrad and 18.06 mrad, respectively) and three different values for β (see table S1), for a total of six combinations. The shift of the bulk plasmon EELS peak with temperature was determined by collecting the EELS maps at room temperature (RT, 20 ºC) and high temperature (HT, 1100 ºC) under all conditions, as shown in figure S2. Correspondingly, the average values of *Ep* in each map and the difference in *Ep* when measuring at RT and at HT are shown in table S1.

| α (mrad) | β (mrad) | *Ep* at RT (eV) | *Ep* at HT (eV) | Difference in *Ep* (eV) |
|---|---|---|---|---|
| 1,90 | 1,90 | 24.94 | 24.77 | - 0.17 |
|  | 4,60 | 25.01 | 24.83 | - 0.18 |
|  | 17,47 | 24.98 | 24.78 | - 0.20 |
| 18,06 | 9,03 | 25.01 | 24.84 | - 0.17 |
|  | 18,06 | 24.97 | 24.75 | - 0.22 |
|  | 69,84 | 24.95 | 24.75 | - 0.20 |

*Table S1: Different α and β conditions, and the corresponded average plasmon energy from each Ep map in figure S2. The last column shows the difference in Ep between RT and HT.*

In theory, the estimated shift of the W bulk plasmon energy with temperature is around 0.146 meV/ºC, corresponding to 0.16 eV shift from 20ºC to 1100ºC. Upon comparing all results shown in the last column in table S1, a good agreement with theory can be seen, within a range of 0.06 eV. F

Here, α = 1.9 mrad and β = 4.60 mrad were chosen as optimal condition for our experiments for the following reasons. First, the lower α (= 1.9 mrad) was selected for a high signal-to-noise ratio (SNR) in the collected EELS signal to decrease the beam time on the sample (i.e. minimize potential beam damage effects). Secondly, diffraction effect (i.e. the presence of ED spots) should ideally be excluded from the EELS signal as it complicates determining the location of ZLP. To include only the centre beam (direct electron beam) and simultaneously achieving a high SNR, we choose to collect the EELS with $\beta$ = 4.60 mrad (i.e. larger than centre beam with $\alpha$ = 1.9 mrad) but excluding all diffracted beams.

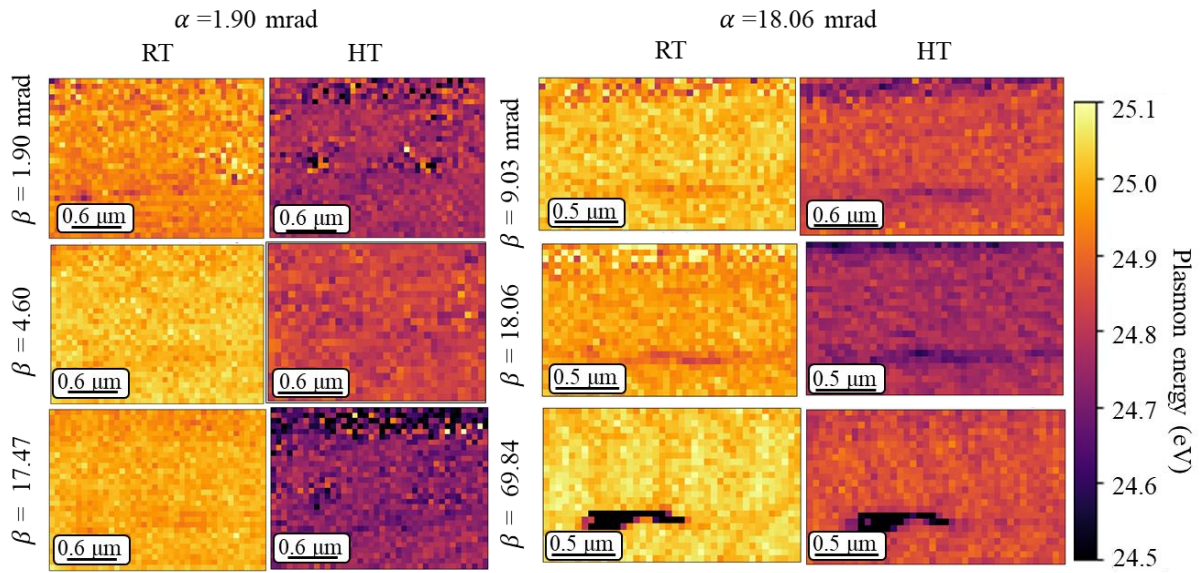

*Figure S2: Bulk plasmon energy (Ep) maps using different α and β at RT and 1100 °C (HT). The left 6 figures using α=1.9 mrad, and from top to bottom row the increase from 1.9, 4.6, 17.47 mrad. The right 6 figures using α=18.06 mrad, and from top to bottom row the β increase from 9.03, 18.06, 69.84 mrad. The colorbar indicates the plasmon energy from 24.5 eV to 25.1 eV. For conditions with α=18.06 mrad, tthe regions where eventually a hole got formed (visible in two bottom right images) where excluded in the calculation of the average Ep stated in table S1.*

### SI.3 Correction for region of interest (ROI) between acquisitions

After a temperature change, especially at the higher temperatures close to 1000°C, slight drifting and bulging of the heating chip occurs causing a shift and a defocusing of the ROI. This has been corrected by translating the ROI to the centre of the screen as well as adjusting the z-height and slightly the focus (kept between – 6 μm and + 6 μm). The mapping window would also get realigned in order to map the same region at each set temperature. Lastly before acquiring a STEM-EELS map an overview ADF image was taken showing the lamella as well as the mapping window over the ROI.

### SI.4 Curve fitting of ZLP and bulk plasmon peak in EELS:

### SI.4.1 Evaluation of curve fitting quality based on χ2-test

1. **Fitting of ZLP with Voigt and Gaussian function.**

In the case of the ZLP, we compared the quality of fitting when using a Voigt and a Gaussian function at 20 ºC. In both cases, the fitting window for ZLP was chosen to be from - 0.5 eV to 0.5 eV. The maps of reduced $\chi^2$ are shown in figure S3. The reduced $\chi^2$ values for the Voigt fit are around 5 to 35 (in figure S3a), and for the Gaussian fitting in the order of $10^4$ (shown in figure S3b). Thus the Voigt fit works best.

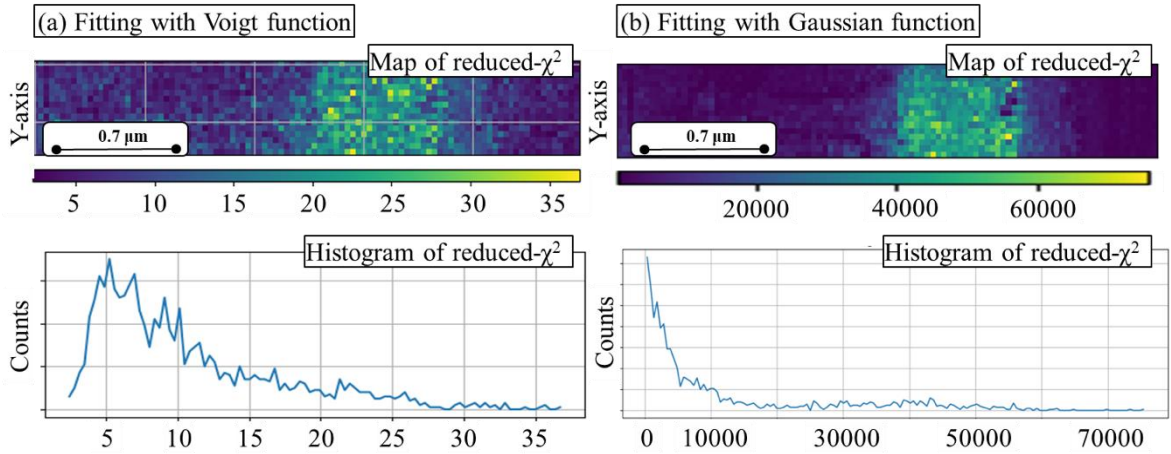

*Figure S3: Maps and histograms of reduced χ2 values for comparing the quality of ZLP curve fitting using a Voigt fit (a) and Gaussian fit (b).*

### 2. Fitting of W bulk plasmon peak with Lorentzian function and Johnson's Su function.

Regarding the fitting of the W bulk plasmon peak, we compared a fitting a Lorentzian function to fitting a Johnson's Su function. The fitting window was taken from 17.5 eV to 30.5 eV.

All maps of reduced $\chi^2$ are shown in figure S4. The values of reduced $\chi^2$ in fitting with Johnson's Su function are around 3 to 5 (in figure S4a), and for the Lorentzian fitting are at the order of $10^3$ to $10^4$ (as shown in figure S4b). Apparently, the Johnson's Su function gives low reduced $\chi^2$ values, indicating the Johnson's Su function fits the experimental data best.

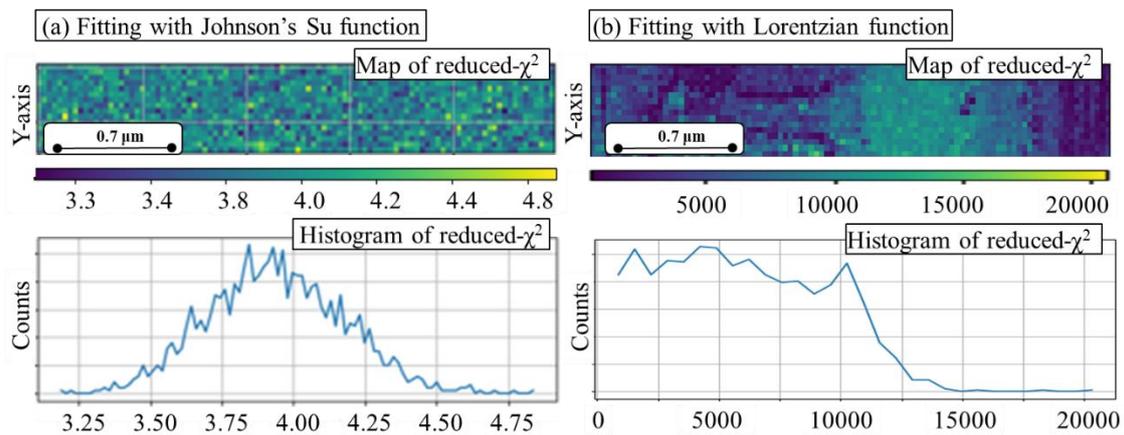

*Figure S4: Maps and histograms of reduced χ2 values for comparing the quality of W bulk plasmon peak curve fitting using the Johnson's Su function (a) and Lorentzian function (b).*

According to the above comparison, we selected the Voigt function for ZLP curve fitting and Johonson's Su function for W plasmon peak fitting.

***SI.4.2 comparison of bulk plasmon fits at 20 ℃ and 700 ℃ at different probe positions***

Figure S5 shows the Johnson's Su fit of the plasmon peak at different probe positions. The top four figures show the plasmon peak located at the thin sample region while the bottom four figures are taken at thicker sample regions. The curves corresponding to the thin region show to be broader and more asymmetric compared to the ones corresponding to the thick region. However at all times the Johnson's Su curve matches well with the experimental data as further confirmed by the low reduced $\chi^2$ values that are homogeneously distributed in the map in figure S4.a.

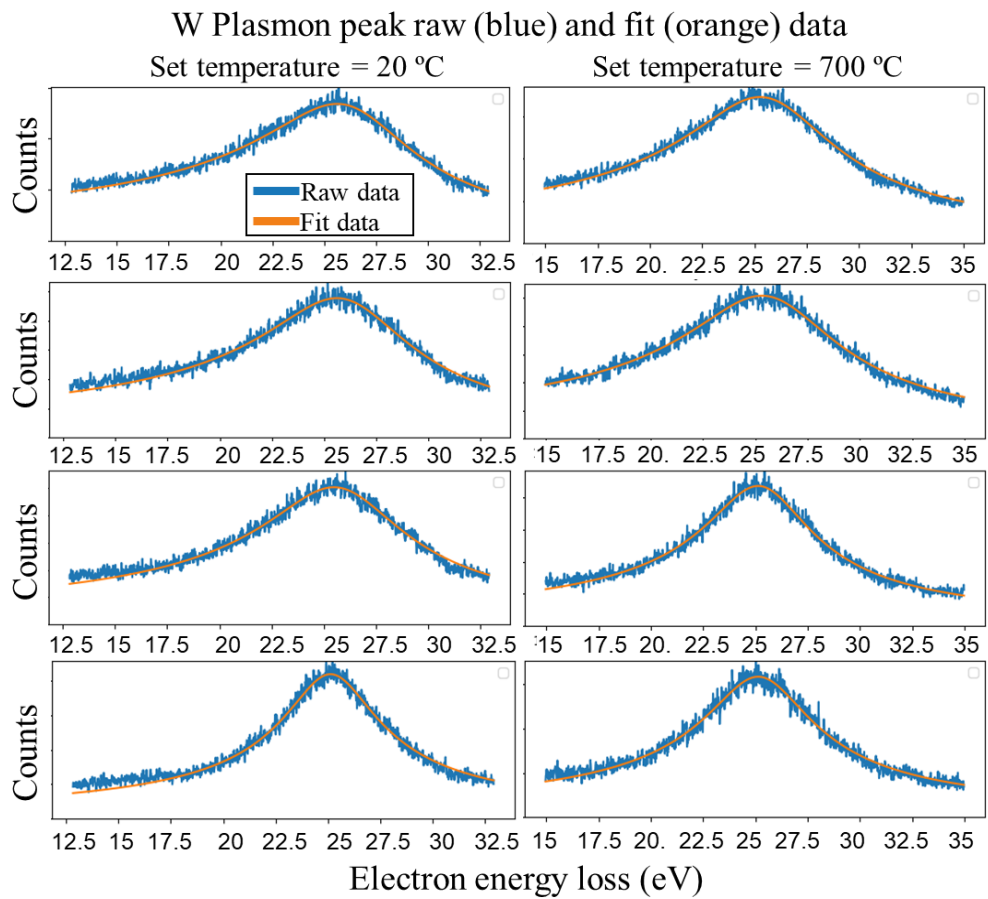

*Figure S5: Raw and the fitted curve of W bulk plasmon peak at different location at 20°C (left column) and 700°C (right column). Top 4 figures are extracted from thin region while bottom 4 from thicker regions.*

### SI.4.3 Ep and FWHM maps from 20 °C to 1000 °C

An overview of the Ep maps and FWHM maps for all the considered set temperatures is given in respectively figure S6 and S7.

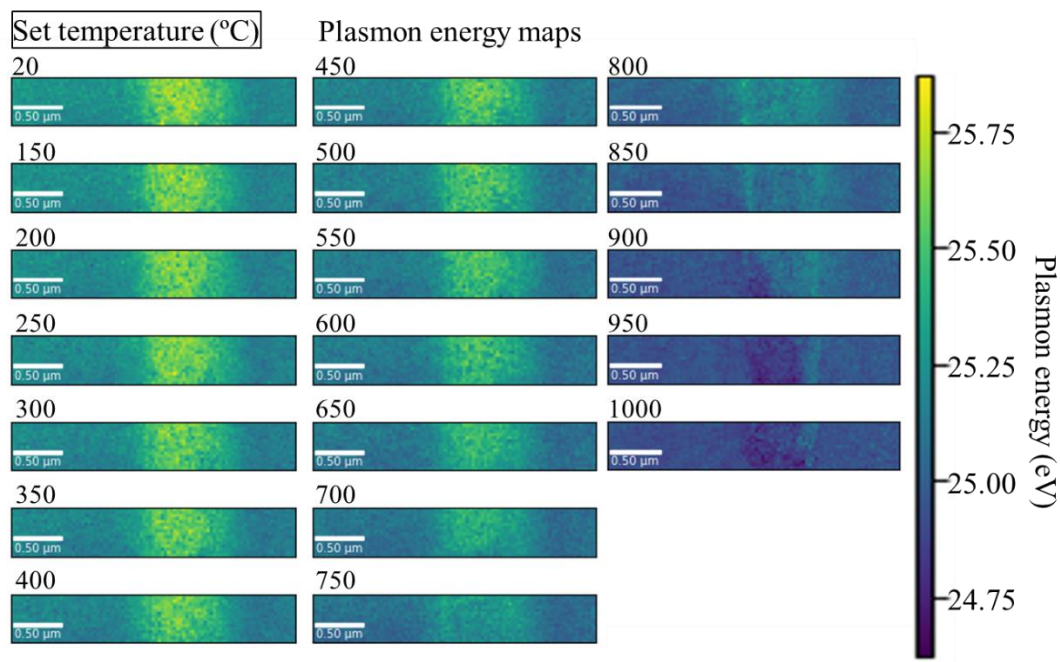

*Figure S6: Maps of Ep over the W-lamella with varying thickness at set temperature from 20°C to 1000ºC. The set temperature in unit of ºC is annotated above each map. The colour bar shows the plasmon energy from 24.6 eV to 25.8 eV.*

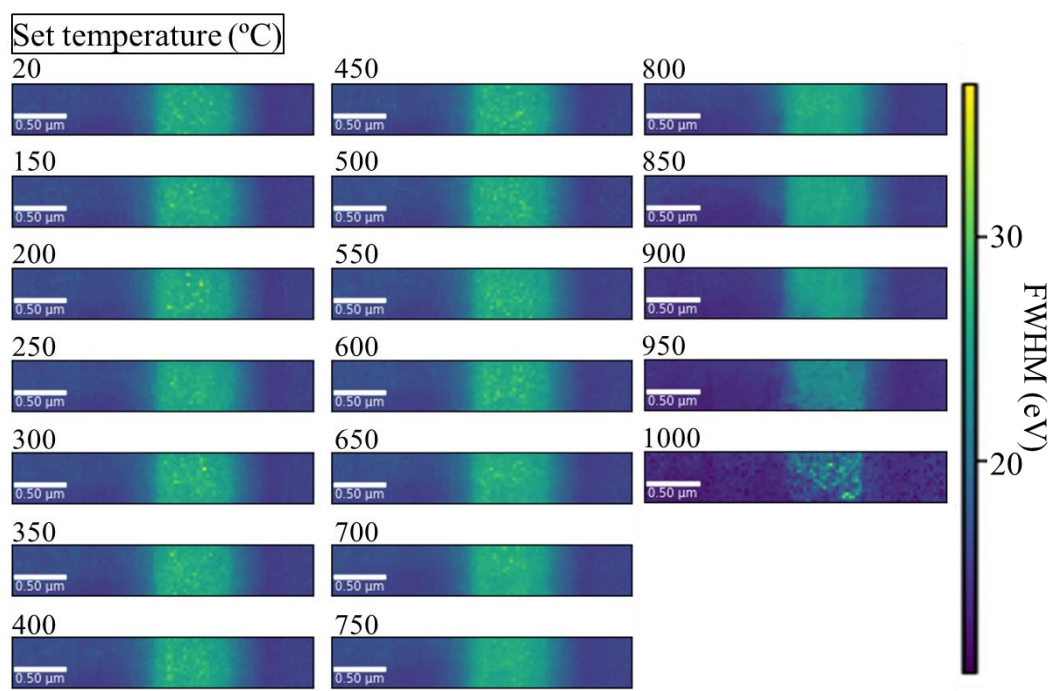

*Figure S7: Maps of the FWHM of plasmon peak over the W-lamella with varying thickness at set temperature from 20°C to 1000ºC. The set temperature in unit of ºC is annotated above each map. The colour bar shows the FWHM of W plasmon from 10 eV to 40 eV.*

### SI.5 Deviating behavior in sample at temperatures above 750 ºC

In the main text, due to some indications of sample morphology changes in the experiments, limiting the applicability of PEET, we only considered experimental data acquired to a maximum temperature of 700ºC. First, annular dark-field (ADF) STEM images shown in figure

S8 indicate that the width of the W-lamella is expanding when increasing the heating temperature from 20ºC to 800ºC. Here, we measured a certain length from the same feature (see dashed line in figure S8) as a measure of the width of the lamella. However, starting at temperatures above 800 ºC, the assigned length (dashed line) starts decreasing. At 20 ºC, the length is 4.00 μm, and expanding to 4.02 μm at 650ºC, 4.03 μm at 700ºC, 4.04 μm at 750ºC and 4.08 μm at 800ºC. Then the length decreases to 4.07 μm at 850ºC and then 4.03 μm at 900ºC, respectively. The error for all the values is less than 0.01 μm (pixel size in the ADF images).

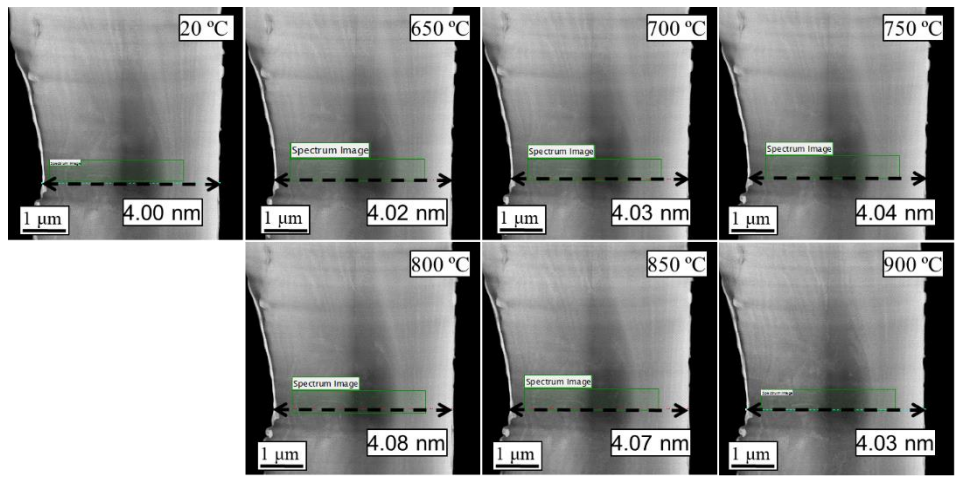

*Figure S8: ADF images at 20°C, 650°C, 700°C, 750°C, 800°C, 900°C. At each image, the length of the arrow has measured from the same feature at left side, and the arrow and length are indicated in the images.*

Secondly, the total transmitted intensity ($I_t$) maps further indicate potential sample morphology changes taking place in the thin region starting at 750°C, as indicated with a red circle in figure S9. This change in scattering behavior suggests a change in microstructure in the sample. The change in morphology and the relative thickness of less than 60 nm could cause deviation from 'normal' thermal expansion of the materials (i.e. limiting the application of PEET) at temperatures above 700 ºC.

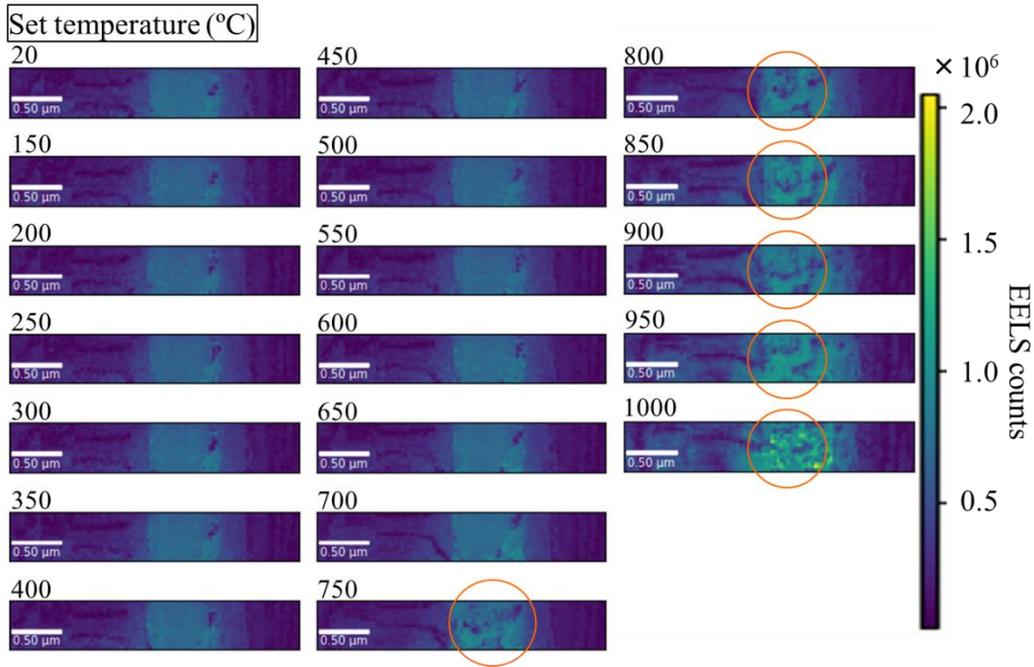

*Figure S9: Maps of the total EELS counts over the W-lamella with varying thickness at set temperature from 20 to 1000 ºC. The set temperature in unit of ºC is annotated above each map. The colour bar shows the total EELS counts from 0 to 2×10$^6$. The red circle indicates the change in EELS counts (at the thin region of the sample).*

Our speculation is that the morphological changes which began at a temperature = 750°C are due to recrystallisation in the material. Although the recrystallisation temperature for bulk W materials is believed to be at 1100°C to 1250°C [2]. However, at the nanoscale (< 50 nm thickness) and in vacuum surrounding this recrystallisation temperature is expected to be much lower, especially when strain is induced from the FIB sample preparation procedure (as indicated in SI.7). *Note: Melting should not yet occur, as the melting temperature of bulk W is 3410°C [3].*

### *SI.6 Accuracy of the COMSOL simulation in comparison to previous studies*

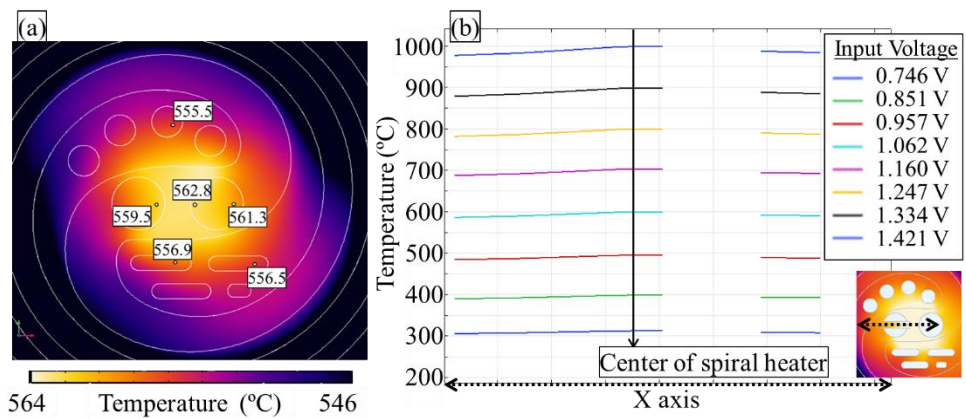

*Figure S10: Finite element modelling of temperature distribution over the MEMS heating chip. (a): Temperature distribution at a set temperature of 550℃. (b) line profiles (along dashed arrow of bottom right subfigure) display the simulated temperatures at the centre of the heating spiral for set input voltages stated in the legend. These input voltages correspond to set (nominal) temperatures from 300°C to 1000°C in steps of 100°C.*

The simulated temperature distribution at the centre of the spiral using COMSOL (see *Methods*), as shown in figure S10a, is in agreement with the simulation at the same set temperature of 550 ºC published by van Omme et al.[4]. More specifically the temperature at the selected points in our simulation, shown in figure S10a, match well with their result from a temperature measurement in an *in-situ* experiment [4]. This result indicates that the considered simulation model of the MEMS heating chip allows accurate simulations of the heat distribution.

In the simulations we considered the temperature at the centre of the heating spiral as the set (nominal) temperature. In order to have a set temperature between 300°C to 1000ºC, an electric voltage between 0.746 V to 1.421 V had to be considered, as shown in figure S10b.

### SI.7 Simulation of stress distribution across the W lamella

Using the same simulation setup as in figure 2 in the main text, we further extracted the stress value along the thickness gradient area, as shown in figure S11. The overall trend shows that a higher stress has been introduced at the area with lower sample thickness. This would support the perspective that the thickness-induced stress could cause the thickness-dependent shift in *Ep*.

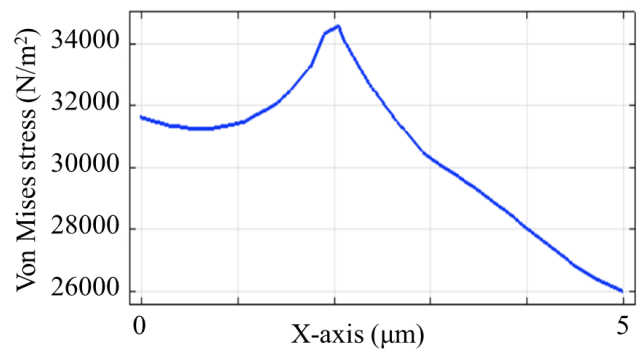

*Figure S11: Line stress profiles at set temperature of 20 ºC along the double-arrow line in figure 2b. Corresponded to thickness profile in figure 2c, the higher stress present at the thinner area.*

### SI.8 Estimation of local sample heating by the electron beam:

The amount of electron beam energy dissipated into the sample can be estimated by the stopping power based on Bethe-Block formula [5,6]. The steady-state temperature difference can be calculated using the heat equation applied to a 2D disk of height ($h$). Let us assume that the beam irradiated area and heat dissipation is confined to within a radius ($r_b$) of the disk's centre and let the sink be located at $r_m$. The temperature difference ($\Delta T$) in the beam irradiated area relative to the sink at room temperature is $\Delta T = \frac{Q}{h}\frac{\ln(r_m/r_b)}{2\pi\kappa}$, where $Q$ is the net heating power (in W) and $\kappa$ is the material's thermal conductivity (in W/m·K). Similar equations for estimating electron beam heating have been derived in previous works [7–9].

In the case of W and a 300 kV electron beam, the stopping power is around 1.292 MeVcm²/g [10]. An elevated electron beam current of 5.5 nA was considered and 1 μs of dwell time. Beam size ($r_b$) is assumed to be 10 nm and the irradiated area ($r_m$) is about 1 μm. Then the net heating power can be estimated at about 1.375 × 10$^{-12}$ W. Assuming a sample thickness of 70 nm and using the thermal conductivity of W of 173 W/m·K, then the temperature increase by electron beam heating can be estimated to be 0.019 ºC, well below the temperature provided by Joule heating from the MEMS heater.